\documentclass{llncs}
\usepackage{graphicx}
\usepackage[margin=1.6in]{geometry}

\begin{document}

\title{Empirical Evaluation of a Thread-Safe Dynamic Range Min-Max Tree using HTM}
\titlerunning{Hamiltonian Mechanics}  
%
\author{Erick Elejalde \and Jose Fuentes-Sep{\'{u}}lveda \and Leo Ferres}
\authorrunning{Erick Elejalde et al.} 
%
\tocauthor{Erick Elejalde, Jose Fuentes-Sep{\'{u}}lveda, Leo Ferres}
\institute{University of Concepcion, Concepcion, Chile,\\
\email{\{eelejalde|jfuentess|lferres\}@udec.cl}}

\maketitle              

\begin{abstract}
Succinct trees, such as wavelet trees and those based on, for instance, range
Min-Max trees (RMMTs), are a family of practical data structures that store information close to their information-theoretic space lower bound.
These structures are often static; meaning that once they are built, nodes cannot be
added, deleted or modified. This read-only property simplifies concurrency. However,
newer versions of these data structures allow for a fair degree of dynamism.
Parallel programming using Hardware Transactional Memory(HTM), has been
available in mainstream microprocessors since a few years ago. One limitation of HTM is still on the size
of each transaction. This is why HTM's use, for the moment, is limited to operations that involve few
memory addresses that need to be updated atomically, or where the level of concurrency
is low. We provide the first available implementation of a concurrent, dynamic RMMT based on HTM, and we compare
empirically how well HTM performs compared to a naive implementation using locks.
We have shown that because of the formal properties of RMMTs, HTM is a good fit for
adding concurrency to otherwise slow lock-based alternatives. We have also shown that HTM performs better than locks when the number of write operations increase, making
it a practical structure to use in several write-intensive contexts.
This is, as far as we know, the only practical implementation of RMMTs thoroughly tested using HTM.
\end{abstract}
\vspace{-4mm}
\section{Introduction}
\vspace{-4mm}
Succinct trees (such as wavelet trees (WTs) \cite{wtfa:gnavarro} and those based on, for instance, range Min-Max trees \cite{Navarro:2014:FFS:2620785.2601073}, to name some of the most common) are a family of practical data structures that store information close to their information-theoretic space lower bound. These structures are often ``static''; meaning that once they are built, nodes cannot be added, deleted or modified. This read-only property simplifies concurrency. However, newer versions of these data structures allow for a fair degree of dynamism \cite{Navarro:2014:FFS:2620785.2601073}. As far as we know, no research has been done on making succinct data structures both dynamic and concurrent, taking advantage of the processing power of multicore architectures.

One problem in making these structures concurrent is that common locks are usually slow. Parallel programming using Hardware Transactional Memory(HTM) \cite{Herlihy:1993:TMA:165123.165164}, has been available in mainstream microprocessors since a few years ago. Perhaps given its novelty, there are still a few restrictions using HTM in practice. One is the limitation on the size of each transaction. The more instructions a transaction involves, the more likely it will fail. This is why HTM's use, for the moment, is limited to operations that involve few memory addresses that need to be updated atomically, or where the level of concurrency is low.
 
Interestingly, the critical regions of dynamic structures such as the range Min-Max tree are small \cite{DBLP:conf/wea/SepulvedaEFS14,Ferres2015}. In the WTs for example, the height of the tree depends on the size of the alphabet, not the text. The range Min-Max tree, in turn, used to index the succinct representation of any tree, the height is constant under the word-RAM model. However, implementing concurrent dynamic trees always poses a challenge. Particularly when any data modification involves the update of the entire path from the root to, at least, one leaf. Since every path shares at least the root with other paths, update operations generate a lot of contention for concurrent accesses, even for read-only queries. Some other methods for allowing concurrency in dynamic trees, like patching the tree, usually need a considerable amount of extra space. The use of extra space is generally not advised whenever succinct structures are required. The alternative is a lock-based implementation. In this case, since every writing operation must reach the leaves, modify it and then traverse back to the root updating all nodes in the path, it is almost unavoidable to use a global lock over the entire data structure to protect threads from leaving the tree in an invalid state.

Our contribution in this paper is two-fold: we provide the first available implementation of a concurrent, dynamic range Min-Max tree based on HTM, and we compare empirically how well HTM performs compared to a naive implementation using locks, given that these two techniques present a similar degree of complexity for the programmer. 
\vspace{-2mm}
\section{Implementation}
\vspace{-4mm}
In these pages we focus on dynamic, concurrent Range Min-Max trees (RMMTs). The RMMT structure has been described at length in \cite{Navarro:2014:FFS:2620785.2601073}, but only theoretically, and its practical behavior is still poorly understood. Concurrency aspects of RMMTs, in turn, have not yet been explored at all. RMMTs may be implemented in different ways. The original paper lists two of them: a balanced binary tree and a B-tree. The binary tree version has been implemented by \cite{Joannou:2012:DST:2366685.2366707}, where the authors use a splay tree to improve the practical performance of some operations. Given that our idea hinges on the control of the height of the tree and the alignment of the nodes in memory, we opted for the alternative route, a B+ tree. In \cite{KarnagelDRLLSL14} the authors evaluate the performance of HTM for a similar indexing structure but with a focus on Data Bases. In our case each node needs to store more information about the excesses but the arity of our tree is smaller. Given the underlying information, the operations over the data structure need to follow a different semantic. For example, insertions and deletions in our data structure have to be made in pairs (we have to remove/insert open-close matching parenthesis corresponding to a node of the underlying tree). We also complement their experiments with a bigger number of threads to investigate the behavior of our data structure under stress conditions.

To encode any tree succinctly, we can use a balanced parentheses (BP) representation, created from a depth-first search on the tree: upon reaching a new node, a bit value is written, say 1, representing an open parenthesis. A bit value 0 is written after visiting all of its children. Since we only need one bit to represent an open or close parenthesis, and every node involves both an open and a close parenthesis, this representation takes $2n$ bits of space. To be able to answer queries, such as navigating to an specific node and deleting or inserting nodes, and make it efficiently, we need to build an index over the BP sequence. Here is where the RMMT comes into play. The BP sequence is partitioned in blocks of S bits and each block is assigned to a leaf of the RMMT. From the leaves to the root, each node represents the range of the BP sequence covered by its subtree. Each node stores the total excess on its range. The excess is the number of parenthesis opened but not yet closed in the range. The excess up to a given position in the sequence represents the depth of the underlying tree at that point. Each node in the RMMT also stores the maximum excess in the range, the minimum excess, how many times this minimum value appears, and finally the number of parenthesis in its range. 

Since the set of HTM instructions that we used (Intel TSX) have a granularity of CPU cache lines (64 bytes for Haswell), this means that one transaction might abort because of false-sharing with another transaction. So we tried to ensure that each transaction's updates affect as few nodes as possible, counting as affected nodes all those whose cache line is modified. To make the tree more HTM friendly, we force each node to take an entire cache line, so modifications on a given node do not affect cache lines of other transactions working in parallel over a different set of nodes. To do this without wasting any memory, we select the k-ary of the tree in a way that the k pointers added to the information needed by the node will complete the 64 bytes of memory of a cache line.

In our implementation, each node of the RMMT may have up to five children. In our 64-bit architecture the five pointers plus the five values of excess mentioned above add up to 60 bytes per node. The other 4 bytes needed to pad the cache line are used to differentiate the internal nodes from the leaves. In the leaves, the five-word size space used by the pointers is used instead to store the actual range of bits of the BP sequence.

To keep the tree balanced, the normal procedure of B-trees is followed. A node is split if it overflows the maximum number of children. If we perform a delete operation and a node is left with less than half the maximum number of children, it tries to ``steal'' some children from an adjacent node, but if the victim node cannot spare any child, both nodes are merged.
\vspace{-2mm}
\section{Experiments}
\vspace{-4mm}
All algorithms were implemented in the C programming language and compiled using GCC 4.9.2 using the -O3 optimization flag. The experiments were carried out on an Intel Xeon E3-1225 v3, with 4 physical cores running at 3.2 GHz. There were only 4 hardware threads running. The computer runs Linux 3.13.0-49-generic, in 64-bit mode. This machine has a per-core L1 and L2 caches of sizes 32KB and 256KB, respectively and a shared L3 cache of 8192KB, with a 12,211,900KB ($\sim$12GB) DDR RAM memory. All alternatives solutions were compared in terms of throughput using the usual high-resolution (nanosecond) C functions in $\langle$time.h$\rangle$.

Our experiments consisted of creating a given number of threads and then let each one runs for an specific amount of time (10 seconds in our set-up) in a loop. In every iteration, each thread has a predefined chance of entering the data structure as a reader or as a writer. For the lock-based version we use a global lock over the entire data structure. In the case of HTM, one transaction includes all operations, from finding the leaf to updating the path to the root. Each transaction, if it aborts, falls back to a lock-protected critical section.

We run the experiments making the transaction to retry in case of an abort from 0 to 2 times before going through the fallback. Although the absolute values change, the relative trend between HTM and Locks keeps the same. Even the commit-fallback ratio stays almost the same.

Besides the critical section, each iteration runs a non-critical section that repeatedly modifies a private vector. The non-critical section always runs for $\frac{1}{100}$ of the total amount of time assigned to the outer loop. We where more interested in the interaction between multiple thread than how many operations a single thread could perform, so we ran experiments varying the number of concurrent threads from 10 to 260.

For the lock version we are using the RW-lock from the Posix {\em pthread} library. Since, in practical applications, an important part of the accesses to this structure are read-only.

We tested our algorithm on different inputs. All of them are succinct representations of a tree. We report here the results for the XML tree of the Wikipedia dump\footnote{http://dumps.wikimedia.org/enwiki/20150112/enwiki-20150112-pages-articles.} (498,753,914 parentheses).

\begin{figure}[t]
  \begin{minipage}{0.48\linewidth}
     \includegraphics[scale=0.38]{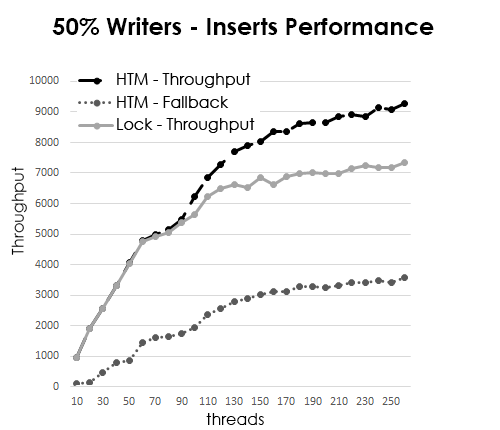}
     \caption{Scaling on HTM and RW-Lock (wikipedia)}
     \label{fig:image1}
  \end{minipage}
  \hfill
  \begin{minipage}{0.5\linewidth}
    \hfill
    \includegraphics[scale=0.38]{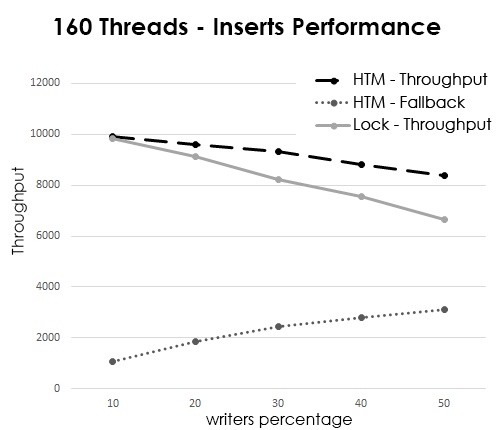}
    \caption{Varying the writing percentage on HTM and RW-Lock (wikipedia)}
    \label{fig:image2}
  \end{minipage}
  \vspace{-4mm}
\end{figure}

Figure \ref{fig:image1} shows that, at the start, both the HTM and lock-based implementation of the RMMT behave similarly (we report the mean over three repetitions of the experiment). Beyond 50 threads, however, HTM shows noticeably higher throughput. As expected, the slowdown in the lock-based version coincides with an increase in HTM fallbacks (slow path). At this point, threads begin to interfere with each other. Even with this increase in fallbacks, the HTM throughput continues growing at a fast pace. Around approximately 120 threads HTM also starts to decelerate as it gets closer to 10000 ops, where our implementation reaches its limit performance.

We have also compared the behavior of both alternatives (HTM and Locks) varying the percentage of writers. Figure \ref{fig:image2} shows these results. For environments with a low percentage of writers the lock-based version behaves comparably to the HTM implementation because of the RW-lock. In other words, a few writers do not generate too much contention. At 50\% of writing operations, HTM throughput exceeds the lock-based implementation by $20\%$ for 160 threads and $26\%$ with 260 threads. Thus, in write-intensive environments, such as many XML documents in the web, the use of HTM represents an important improvement over locks.
\vspace{-2mm}
\section{Conclusions and Future Work}
\vspace{-4mm}

We have shown that because of the formal properties of RMMTs, HTM is a good fit for adding concurrency to otherwise slow lock-based alternatives. We have also shown that HTM performs better than locks when the number of write operations increase, making it a practical structure to use in several write-intensive contexts. This is, as far as we know, the only practical implementation of RMMTs thoroughly tested using HTM.

In the future, we plan to test the behavior of other alternatives to allow concurrency, such as NUMA-based locking. These may well complement the current proposal. 

We could also explore batch processing, grouping a few insertions and deletions in the same BP block in a single transaction. It could be the case that a set of update operations ``cancel'' each other's effect on the excesses counters and prevent having to spread the actualization to higher levels in the RMMT, conflicting with other transactions. 

We also want to experiment changing the data structure to make it more HTM friendly. For instance, storing the length of the segment on a different structure may reduce conflicts. The length value changes frequently, even when all other values remain unchanged.

%
%
\bibliographystyle{IEEEtran}
\bibliography{IEEEabrv,bibtex}

\end{document}